\documentclass[11pt,superscriptaddress,aps,prd,preprint]{revtex4}
\usepackage{graphicx}
\usepackage{amsmath}
\usepackage{slashed}
\usepackage{hyperref}

\usepackage[latin1]{inputenc}

\begin{document}
\title{Nonanalyticity of the nonabelian five-dimensional Chern-Simons term}
\author{J. F. Assun\c c\~ao}
\affiliation{Instituto de F\'\i sica, Universidade Federal de Alagoas, 57072-900, Macei\'o, Alagoas, Brazil}
\affiliation{Departamento de F\'\i sica, Universidade Regional do Cariri, 63180-000, Juazeiro do Norte, Cear\'{a}, Brazil}
\author{J. Furtado}
\affiliation{Centro de Ci\^{e}ncias e Tecnologia, Universidade Federal do Cariri, 63048-080, Juazeiro do Norte, Cear\'{a}, Brazil}
\author{T. Mariz}
\affiliation{Instituto de F\'\i sica, Universidade Federal de Alagoas, 57072-900, Macei\'o, Alagoas, Brazil}
\email{ jfassuncao,job.furtado,tmariz@fis.ufal.br}
\date{\today}
\begin{abstract}
In this work, we study the behavior of the nonabelian five-dimensional Chern-Simons term at finite temperature regime in order to verify the possible nonanalyticity. We employ two methods, a perturbative and a non-perturbative one. No scheme of regularization is needed, and we verify the nonanalyticity of the self-energy of the photon in the origin of momentum space by two conditions that do not commute, namely, the static limit $(k_0=0,\vec k\rightarrow 0)$ and the long wavelength limit $(k_0\rightarrow 0,\vec k= 0)$, while its tensorial structure holds in both limits. 
\end{abstract}

\maketitle
\section{Introduction}

It is well known that in gauge theories with fermions, a topological mass term can be generated dynamically and that, in this case, the effective theory contains a Chern-Simons (CS) term. This term is topological and parity-odd, being introduced and first studied in \cite{Schonfeld:1980,Deser:1982}. It gained attention in $2+1$ dimensions due to its relation with planar phenomena, such as superconductivity and the fractional quantum Hall effect \cite{Wilczek:1982,Wilczek:1983}. Furthermore, in the presence of such a term, the gauge fields become massive and for a non-abelian gauge theory, the quantization of the mass parameter appears as a consequence of the large gauge invariance. While the large gauge invariance mechanism, and consequently the origin of mass quantization, was well understood at zero temperature, at finite temperature scenario it only became possible due to the works \cite{Dunne:1996, Deser:1997}, which showed that only the complete effective action is invariant under large gauge transformations, at least for a specific choice of gauge field background.

The CS term can be generalized to any arbitrary odd-dimension $d=2n+1$ of spacetime. The full form (topological structure) of non-abelian CS term,
\begin{eqnarray}\label{}
Q_{2n-1}(\omega,\Omega)=n\int_{0}^{1}dt P_n(\omega,\Omega_{t}^{n-1}),
\end{eqnarray}
can be obtained from the invariant polynomial $P_{n}(\omega,\Omega^{n-1})=Tr\,\omega\,\Omega^{n-1}$ in one dimension higher, constructed from the connection $\omega$ and the curvature two-form $\Omega=\Omega^{a}\tau^{a}$, being $\tau_{a}$ the generators of some Lie group and $\Omega_{t}=t\Omega+(t^2-t)\omega^2$ \cite{Bertlmann:1996}. The connection $\omega$ corresponds to different quantities in different theories.  

Studies in five-dimensional quantum field theories have been attracted much attention recently. A correspondence between a six-dimensional field theory compactified on a circle and a five-dimensional field theory, propagating both massive and massless degrees of freedom, is, basically, the origin of the interest of studying five-dimensional CS theories \cite{Grimm:2013}. It is expected that the five-dimensional CS coupling reveals information on higher dimensional anomalies.

Beyond the anomaly cancelation, the CS five-form was used to construct a topological gauge theory of gravity in five dimensions \cite{Chamseddine}. In this context of a gauge theory of gravity, the CS form has already been used by Witten \cite{Witten2} to construct a theory of gravity that is not only renormalizable but finite as well.  

At finite temperature regime, the one-loop radiative corrections are nonanalytic functions in the origin of momentum space ($k_{\mu} = 0$). Thus, as due to the choice of a specific frame by the thermal bath the radiative corrections, in general, has different dependencies of $k_0$ and $\vec{k}$, the two conditions ($k_0 = 0,\vec k\rightarrow 0$) and ($k_0\rightarrow 0,\vec k = 0$) do not commute. In fact, this has been shown in the case of Lorentz-violating QED \cite{Assuncao:2016fko}, three-dimensional QED \cite{KaoYang:1993}, hot QCD \cite{Gross:1980br,Weldon:1982aq,Frenkel:1989br}, self-interacting scalars \cite{Das:1997}, and Maxwell-Chern-Simons-Higgs model \cite{Alves:2001nm}. The first condition ($k_0 = 0,\vec k\rightarrow 0$) is sometimes referred as the ``static" limit, while the other condition ($k_0\rightarrow 0,\vec k = 0$) is the ``long wavelength" limit. In the physical context, these conditions are related to the Debye and plasmon mass responsible for the screening of the gauge field (the damping of the gauge field caused by the presence of the thermal virtual pairs), respectively.

In this paper, we study the behavior of the five-dimensional CS coefficient in order to verify a possible nonanalyticity. Hence we analyze if there is any change in the topological structure of the induced CS term when the physical system is placed in contact with a thermal bath. In section II, we single out the five-dimensional CS term through derivative expansion, and the finite temperature contribution is achieved in the imaginary time formalism using the Ford expression \cite{Ford:1979ds} to compute the sum over the Matsubara frequencies. In section III, we calculate the triangle, box, and pentagon graphs individually, in a non-perturbative approach, to analyze a possible nonanalytical behavior of the five-dimensional CS coefficient. Finally, in section IV, we present a brief summary of our results.

In this paper, we are working with five-dimensional gamma matrices $\gamma^{\alpha}$, which are complex-valued $4\times4$ matrices satisfying the anticommutation relation
\begin{equation}
\{\gamma^{\mu},\gamma^{\nu}\}=2g^{\mu\nu},
\end{equation}
where  $g^{\mu\nu}=\mathrm{diag}(+,-,-,-,-)$ is the Minkowski metric. The trace calculation of such gamma matrices obey the following rules:
\begin{subequations}\label{trgamma}
\begin{eqnarray}
tr(\gamma^{\mu})&=&0,\\
tr(\gamma^{\mu}\gamma^\nu\gamma^\lambda)&=&0,\\
tr(\gamma^{\mu}\gamma^{\nu}\gamma^{\lambda}\gamma^{\rho}\gamma^{\alpha})&=&4i\epsilon^{\mu\nu\lambda\rho\alpha},\\
\nonumber tr(\gamma^{\mu}\gamma^{\nu}\gamma^{\lambda}\gamma^{\rho}\gamma^{\alpha}\gamma^{\beta}\gamma^{\sigma})&=&4i \left(g^{\beta\alpha} \epsilon^{\lambda  \mu  \nu  \rho  \sigma }+g^{\rho\nu} \epsilon^{\alpha  \beta  \lambda  \mu  \sigma }-g^{\rho\mu} \epsilon^{\alpha  \beta  \lambda  \nu  \sigma }+g^{\rho\lambda} \epsilon^{\alpha  \beta  \mu  \nu  \sigma }+g^{\rho\beta} \epsilon^{\alpha  \lambda  \mu  \nu  \sigma }+\right.\\
\nonumber&&\left.-g^{\rho\alpha} \epsilon^{\beta  \lambda  \mu  \nu  \sigma}+g^{\nu\mu} \epsilon^{\alpha  \beta  \lambda  \rho  \sigma }+g^{\nu\lambda} \epsilon^{\alpha  \beta  \mu  \rho  \sigma}-g^{\nu\beta } \epsilon^{\alpha  \lambda  \mu  \rho  \sigma}+g^{\nu\alpha} \epsilon^{\beta  \lambda  \mu\rho\sigma}+\right.\\
&&\left.-g^{\mu\lambda} \epsilon^{\alpha  \beta  \nu  \rho  \sigma }+g^{\mu\beta} \epsilon^{\alpha  \lambda  \nu  \rho  \sigma}-g^{\mu\alpha} \epsilon^{\beta  \lambda  \nu  \rho  \sigma}-g^{\lambda\beta} \epsilon^{\alpha  \mu  \nu  \rho  \sigma }+g^{\lambda\alpha} \epsilon^{\beta  \mu  \nu  \rho  \sigma }\right),
\end{eqnarray}
\end{subequations}
and so on.

\section{The One-loop Induced five-dimension Chern-Simons Term}\label{1Loop}

Initially, we are interested in studying the radiative induction of the five-dimensional CS term. For this, the starting point of our model is the usual fermionic sector of the QED, whose Lagrangian is given by
\begin{equation}
\mathcal{L}=\bar\psi(i\slashed{\partial}-m-e\slashed{A})\psi.
\end{equation}
Consequently, the corresponding generating functional is written as
\begin{equation}
Z[A_{\mu}]=\int D\bar\psi D\psi e^{\int d^5x\mathcal{L}}=e^{iS_{eff}},
\end{equation}
and the integration over the fermions gives us the one-loop effective action
\begin{equation}\label{EffecAction}
S_{eff}=-i\, \mathrm{Tr}\ln(\slashed{p}-m-e\slashed{A}).
\end{equation}
Here, $\mathrm{Tr}$ consists in the trace over the Dirac matrices and over the space coordinates. 

In order to obtain the non-abelian CS term, we must select cubic, quartic, and quintic contributions in $A_{\mu}$. To achieve this, we rewrite the expression (\ref{EffecAction}) as
\begin{equation}
S_{eff}=S_{eff}^{(0)}+\sum_{n=1}^{\infty}S_{eff}^{(n)},
\end{equation}
where $S_{eff}^{(0)}=-i\, \mathrm{Tr}\ln(\slashed{p}-m)$ and
\begin{equation}\label{Seffn}
S_{eff}^{(n)}=\frac{i}{n}\, \mathrm{Tr}\left[S(p)e\slashed{A}\right]^{n},
\end{equation}
with $S(p)=(\slashed{p}-m)^{-1}$. It is important to notice that, as it was first noted in ref \cite{Das1}, there is no nonanaliticity present if the loop involves distinct masses in the propagators.

\begin{figure}[!h]
\begin{center}\includegraphics[scale=0.6]{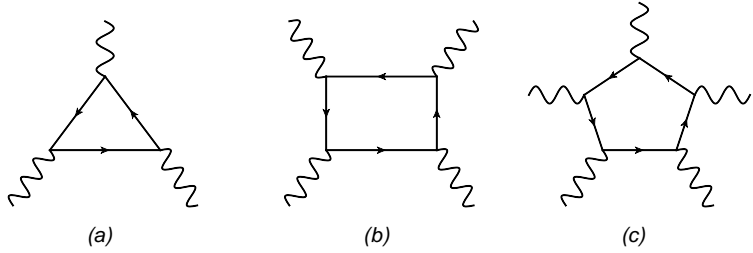}
\caption{Feynman graphs}\label{fig1}
\end{center}
\end{figure}

At this point, we have to consider $n=3,4,5$ in the expression (\ref{Seffn}), evaluate the trace over the space coordinate considering the commutation relation $A_{\mu}(x)S(p)=S(p-i\partial)A_{\mu}(x)$, and  use the completeness relation of the momentum space. Then, for $n=3$, we get
\begin{equation}
S_{eff}^{(3)}=\frac{ie^3}{3}tr\int d^5x\int\frac{d^5p}{(2\pi)^5}S(p)\gamma^{\mu}S(p_1)\gamma^{\nu}S(p_{12})\gamma^{\lambda}A_{\mu}A_{\nu}A_{\lambda},
\end{equation}
where $p_1=p-i\partial_1$,  $p_{12}=p-i\partial_1-i\partial_2$, and so on. Performing a derivative expansion in the external momentum and selecting the terms that will contribute to CS, effective action $S_{eff}^{(3)}$ becomes $S_{CS}^{(3)}$, which is written as
\begin{equation}\label{essa1}
S_{CS}^{(3)}=-\frac{ie^3}{3}tr\int d^5x\int\frac{d^5p}{(2\pi)^5}S(p)\gamma^{\mu}S(p)\gamma^{\nu}S(p)\gamma^{\lambda}S(p)\gamma^{\rho}S(p)\gamma^{\sigma}(\partial_{\nu}A_{\mu})(\partial_{\rho}A_{\lambda})A_{\sigma}.
\end{equation}
Analogously, the contributions for $n=4$ and $n=5$ yield
\begin{equation}\label{essa2}
S_{CS}^{(4)}=-\frac{e^4}{2}tr\int d^5x\int\frac{d^5p}{(2\pi)^5}S(p)\gamma^{\mu}S(p)\gamma^{\nu}S(p)\gamma^{\lambda}S(p)\gamma^{\rho}S(p)\gamma^{\sigma}(\partial_{\rho}A_{\mu})A_{\nu}A_{\lambda}A_{\sigma},
\end{equation}
\begin{equation}\label{essa3}
S_{CS}^{(5)}=\frac{ie^5}{5}tr\int d^5x\int\frac{d^5p}{(2\pi)^5}S(p)\gamma^{\mu}S(p)\gamma^{\nu}S(p)\gamma^{\lambda}S(p)\gamma^{\rho}S(p)\gamma^{\sigma}A_{\mu}A_{\nu}A_{\lambda}A_{\rho}A_{\sigma}.
\end{equation}
In order to obtain the equations~(\ref{essa1}) and (\ref{essa2}), we have used the cyclic property of the trace and the solution
\begin{equation}\label{Pro}
tr\int\frac{d^5p}{(2\pi)^5}S(p)\gamma^{\mu}S(p)\gamma^{\nu}S(p)\gamma^{\lambda}S(p)\gamma^{\rho}S(p)\gamma^{\sigma}=\alpha\epsilon^{\mu\nu\lambda\rho\sigma},
\end{equation}
where $\alpha$ is a constant. These expressions (\ref{essa1}), (\ref{essa2}), and (\ref{essa3}) are presented in Fig.~\ref{fig1}, respectively.

Therefore, we can construct a nonabelian five-dimensional CS structure as follows:
\begin{eqnarray}\label{CStopology}
\nonumber S_{CS}&=&\int d^5x\int\frac{d^5p}{(2\pi)^5}\,\mathrm{tr}\,S(p)\gamma^{\mu}S(p)\gamma^{\nu}S(p)\gamma^{\lambda}S(p)\gamma^{\rho}S(p)\gamma^{\sigma}\times\\
&&\times\frac{i}{3}[-e^3(\partial_{\nu}A_{\mu})(\partial_{\rho}A_{\lambda})A_{\sigma}+ie^4\frac{3}{2}(\partial_{\rho}A_{\mu})A_{\nu}A_{\lambda}A_{\sigma}+e^5\frac{3}{5}A_{\mu}A_{\nu}A_{\lambda}A_{\rho}A_{\sigma}].
\end{eqnarray}
Note that, despite the logarithm divergence associated with the coefficient of the five-dimensional CS term, it is not necessary to use any kind of regularization because the divergent terms do not contribute to it. This is a general feature of odd dimension theories. Besides, the implementation of temperature effects only affects the coefficient leaving the topological structure unchanged.

The non-abelian five-dimensional CS action, as it appears in equation~(\ref{CStopology}), has this form because it is written in the fundamental representation, in which $A_{\mu}(x)=A_{\mu}^{a}(x)\tau^a$ and $\tau^a$ are the hermitian generators of the Lie Group $SO(1,5)$ satisfying the commutation relations
\begin{equation}
[\tau^a,\tau^b]=if^{abc}\tau^c,
\end{equation}
where $f^{abc}$ is the structure constant of the group. In the adjoint representation, the CS five-form is given by
\begin{equation}
\omega_5=tr\int d^5x\left(A(dA)^2+\frac{3}{2}A^3(dA)+\frac{3}{5}A^5\right).
\end{equation}
Due to the common integral appearing in all three terms of the five-dimensional CS action (\ref{CStopology}), in order to take into account finite temperature effects, we will work uniquely with its tensorial coefficient
\begin{eqnarray}\label{Tof}
\nonumber S^{\mu\nu\lambda\rho\sigma}&=&\frac{i}{3}\mathrm{tr}\int\frac{d^5p}{(2\pi)^5}\frac{(\slashed{p}+m)\gamma^{\mu}(\slashed{p}+m)\gamma^{\nu}(\slashed{p}+m)\gamma^{\lambda}(\slashed{p}+m)\gamma^{\rho}(\slashed{p}+m)\gamma^{\sigma}}{(p^2-m^2)^5}\\
&=&\frac{i}{3}\, S\, \epsilon^{\mu\nu\lambda\rho\sigma}.
\end{eqnarray}

In calculating the trace of the above numerator, we must consider only the terms that contribute to the CS term, i.e., those with an odd power of mass, since such terms are parity-odd, as required. Then, we have
\begin{eqnarray}\label{Tof}
\nonumber S^{\mu\nu\lambda\rho\sigma}&=&\frac{i}{3}\int\frac{d^5p}{(2\pi)^5}\frac{4i m^5-8ip^2 m^3+4ip^4 m}{(p^2-m^2)^5}\epsilon^{\mu\nu\lambda\rho\sigma}\\
&=&\frac{i}{3}\int\frac{d^5p}{(2\pi)^5}\frac{4im}{(p^2-m^2)^3} \epsilon^{\mu\nu\lambda\rho\sigma}.
\end{eqnarray}
where the scalar coefficient $S$ takes the form
\begin{equation}\label{m1}
S=\int\frac{d^5p}{(2\pi)^5}\frac{4im}{(p^2-m^2)^3}.
\end{equation}

To implement the finite temperature effects, we change from Minkowski to Euclidean space and split the internal momentum $p_\mu$ into its spatial and temporal components. For this, we must perform the following procedure: $g^{\mu\nu}\to-\delta^{\mu\nu}$, i.e., $p^2\to -\delta^{\mu\nu}p^\mu p^\nu=-p^2$, as well as 
\begin{equation}\label{}
\int \frac{d^5 p}{(2\pi)^5} \to \int \frac{d^4\vec p}{(2\pi)^4} \,i\int \frac{dp_0}{2\pi}.
\end{equation}
Also, let us assume from now on that the system is in thermal equilibrium with a temperature $T=\beta^{-1}$, so that the antiperiodic (periodic) boundary conditions for fermions (bosons) lead to discrete values $p_0=(2n+1)\frac{\pi}{\beta}$ and $k_0=\frac{2\pi l}{\beta}$, with $n$ and $l$ being integers. Thus, with ${\textstyle\int}\frac{dp_0}{2\pi}\rightarrow \frac{1}{\beta}{\textstyle\sum}_n$, and considering the adimensional variables $\xi_{p}=\frac{\beta\omega_{p}}{2\pi}$ and $\xi=\frac{\beta m}{2\pi}$, where $\omega_{p}=\sqrt{\vec p^2+m^2}$, we get
\begin{eqnarray}\label{E1}
\nonumber S&=&\frac{1}{\beta}\sum_n \int\frac{d^4\vec p}{(2\pi)^4}\frac{4m}{(p^2+m^2)^3}\\
&=&\frac{\xi }{2 (2\pi)^3}\sum_n \left[(n+1/2)^2+ \xi ^2\right]^{-1}.
\end{eqnarray}
We complete our calculation by computing the sum over the Matsubara frequencies using the Ford expression \cite{Ford:1979ds}, given by 
\begin{equation}
\sum_n[(n+b)^2+a^2]^{-\lambda}=\frac{\sqrt{\pi}\Gamma(\lambda-1/2)}{\Gamma(\lambda)(a^2)^{\lambda-1/2}}+4\sin(\pi\lambda)f_\lambda(a,b)
\end{equation}
where
\begin{equation}
f_\lambda(a,b)=\int_{|a|}^{\infty}\frac{dz}{(z^2-a^2)^\lambda}Re\left(\frac{1}{e^{2\pi(z+ib)}-1}\right),
\end{equation}
which is valid for $Re\,\,\lambda<1$. In our case, as we can see from equation~(\ref{E1}), $\lambda=1$, and hence it is out of the range of validity. To deal with this, i.e., to decrease the value of $\lambda$, we use the recurrence relation
\begin{equation}
f_\lambda(a,b)=-\frac{1}{2a^2}\frac{2\lambda-3}{\lambda-1}f_{\lambda-1}(a,b)-\frac{1}{4a^2}\frac{1}{(\lambda-2)(\lambda-1)}\frac{\partial^2}{\partial b^2}f_{\lambda-2}(a,b).
\end{equation} 
Therefore, the CS scalar coefficient is found to be
\begin{equation}\label{EDT0}
S(\xi)=\frac{\tanh(\pi  \xi )}{16 \pi ^2}.
\end{equation}
Our result at zero temperature is in agreement, e.g., with \cite{Grimm,Witten}. A naive procedure was used at finite temperature in \cite{Sisakian:1997} and the same result was found.

Another interesting feature worth commenting on is that, in a perturbative approach, it is not possible to analyze if there is a nonanalytical behavior of the coefficient. And therefore, in the next section, we will perform a non-perturbative calculation to verify if there is nonanalyticity in the five-dimensional CS coefficient.  

\section{Nonanalyticity of the five-dimensional Chern-Simons Term}\label{Non}

\subsection{Triangle Diagram}\label{Tri}

Now, let us analyze the nonanalyticity present in contributing to the CS term from the three-point function at the finite temperature regime. As previously noted in equation~(\ref{Tof}), only linear terms in the mass contribute to the CS term, so that we get $\Pi^{\mu\nu\lambda} \rightarrow \Pi_{CS}^{\mu\nu\lambda}$, i.e., the convergent integral
\begin{eqnarray}\label{C1}
\Pi_{{\footnotesize CS}}^{\mu\nu\lambda}(k_{1},k_{2}) &=& ie^{3} \frac{i}{\beta}\sum_n\int\frac{d^4 \vec p}{(2\pi)^4}\frac{4im\epsilon^{\mu\nu\lambda\alpha\beta}k_{1\alpha}k_{2\beta}}{\left(p^2+m^2\right) \left(p_{1}^{2}+m^2\right) \left(p_{12}^2+m^2\right)}\nonumber\\
&=& e^3\,S(k_1,k_2)\,\epsilon^{\mu\nu\lambda\alpha\beta}k_{1\alpha}k_{2\beta},
\end{eqnarray}
whose coefficient $S(k_1,k_2)$ is in Euclidean space, with $p_{1}^{\mu}=p^{\mu}-k_{1}^{\mu}$ and $p_{12}^{\mu}=p^{\mu}-k_{12}^{\mu}$. Since the topological structure is independent of the internal momentum and directly obtained from the trace evaluation, let us take only its coefficient $S(k_1,k_2)$ to our analysis. Thus, after performing the frequency sum, we are left with the evaluation of integrals
\begin{eqnarray}\label{S}
S \!\!&=&\!\! \beta^4\!\!\int\!\!\frac{d^4 \vec p}{(2\pi)^8}\left\{\!\!\frac{\xi\tanh\left(\pi\xi_p\right)}{\xi_p\left[\xi_{p_1}^2+\left(l_1+i \xi_{p}\right){}^2\right]\left[\xi_{p_{12}}^2+\left(l_{2}+i \xi_{p}\right){}^2\right]}+\frac{\xi\tanh\left(\pi\xi_p\right)}{\xi_p\left[\xi_{p_1}^2+\left(l_1-i \xi_{p}\right){}^2\right]\left[\xi_{p_{12}}^2+\left(l_{2}-i \xi_{p}\right){}^2\right]}\right.\nonumber\\
 && \left.+\frac{\xi\tanh \left(\pi \xi_{p_{12}}\right)}{\xi_{p_{12}}\left[\xi_p^2+\left(l_{12}+i \xi_{p_{12}}\right){}^2\right] \left[\xi_{p_1}^2+\left(l_{2}+i \xi_{p_{12}}\right){}^2\right]}+\frac{\xi\tanh \left(\pi \xi_{p_{12}}\right)}{\xi_{p_{12}}\left[\xi_p^2+\left(l_{12}-i \xi_{p_{12}}\right){}^2\right] \left[\xi_{p_1}^2+\left(l_{2}-i \xi_{p_{12}}\right){}^2\right]}\right.\nonumber\\
&&\left.+\frac{\xi\tanh \left(\pi \xi_{p_1}\right)}{\xi_{p_1}\left[\xi_p^2+\left(l_1+i \xi_{p_1}\right){}^2\right] \left[\xi_{p_{12}}^2+\left(l_{2}-i \xi_{p_1}\right){}^2\right]}+\frac{\xi\tanh \left(\pi \xi_{p_1}\right)}{\xi_{p_1}\left[\xi_p^2+\left(l_1-i \xi_{p_1}\right){}^2\right] \left[\xi_{p_{12}}^2+\left(l_{2}+i \xi_{p_1}\right){}^2\right]}\right\}, \nonumber\\
\end{eqnarray}
% -\frac{\beta ^5 \left(-\frac{2 \tanh \left(\pi  \xi _k\right) \xi _{k+p_1} \xi _{k-p_2} \left(4 l_2 l_1 \xi _k^2+l_1^2 \left(-\xi _k^2+\xi _{k-p_2}^2+l_2^2\right)-\left(\xi _k^2-\xi _{k+p_1}^2\right) \left(-\xi _k^2+\xi _{k-p_2}^2+l_2^2\right)\right)}{\left(2 l_1^2 \left(\xi _k^2+\xi _{k+p_1}^2\right)+\left(\xi _k^2-\xi _{k+p_1}^2\right){}^2+l_1^4\right) \left(2 l_2^2 \left(\xi _k^2+\xi _{k-p_2}^2\right)+\left(\xi _k^2-\xi _{k-p_2}^2\right){}^2+l_2^4\right)}
% -\frac{\xi _k \xi _{k-p_2} \tanh \left(\pi  \xi _{k+p_1}\right)}{\left(\xi _k^2+\left(l_1+i \xi _{k+p_1}\right){}^2\right) \left(\xi _{k-p_2}^2+\left(i \xi _{k+p_1}+l_1+l_2\right){}^2\right)}
% -\frac{\xi _k \xi _{k-p_2} \tanh \left(\pi  \xi _{k+p_1}\right)}{\left(\xi _k^2+\left(l_1-i \xi _{k+p_1}\right){}^2\right) \left(\xi _{k-p_2}^2+\left(-i \xi _{k+p_1}+l_1+l_2\right){}^2\right)}
% -\frac{\xi _k \xi _{k+p_1} \tanh \left(\pi  \xi _{k-p_2}\right)}{\left(\xi _k^2+\left(l_2+i \xi _{k-p_2}\right){}^2\right) \left(\xi _{k+p_1}^2+\left(i \xi _{k-p_2}+l_1+l_2\right){}^2\right)}
% -\frac{\xi _k \xi _{k+p_1} \tanh \left(\pi  \xi _{k-p_2}\right)}{\left(\xi _k^2+\left(l_2-i \xi _{k-p_2}\right){}^2\right) \left(\xi _{k+p_1}^2+\left(-i \xi _{k-p_2}+l_1+l_2\right){}^2\right)}
% \right)}{128 \pi ^5 \xi _k \xi _{k+p_1} \xi _{k-p_2}}
where we have again used the adimensional variables $\xi_{p_{12}}=\frac{\beta\omega_{p_{12}}}{2\pi}$, with $\omega_{p_{12}}=\sqrt{\vec{p}_{12}^2+m^2}$, and so on, and considered $k_0=\frac{2\pi l}{\beta}$ and $\tan[\pi(l\pm i\xi_{1})]=\pm i\tanh(\pi\xi_{1})$.

At this point, we take the limits on the external momenta in equation~(\ref{S}). There are four considerations, namely, the double static limit, the double long wavelength limit, and two mixed limits. As previously mentioned, in the static limit, we put $k_0=0$ $(l=0)$, and then take the limit $\vec k\rightarrow 0$, unlike, in the long wavelength limit, we put $\vec k=0$, and then take the limit $k_0\rightarrow 0$ $(l\rightarrow 0)$. We will represent the static and long wavelength limits, e.g., as $[k_1]=(k_{10}=0,\vec k_{1}\rightarrow 0)$ and $\{k_1\}=(k_{10}\rightarrow 0,\vec k_{1}=0)$, respectively. For the double static limit, we have
% In the follows expressions, we will use	square brackets $[k]$ and braces $\{k\}$ to represent the static limit and the long wavelength limit to taken about the external momentum $k$, respectively. For the static limit,
% It's interesting that, the expression in the static limit contains the expression in the long wavelength limit, as we will see below. 
% It's interesting that the double static limit contains the crossed limits and so does to the crossed limits and the double long wavelength limit. So, let us consider the double static limit $([k_{1}],[k_{2}])=(k_{01}=0,\vec{k}_{1}\rightarrow 0,k_{02}=0,\vec{k}_{2}\rightarrow 0)$ about the equation~(\ref{S}),
% \{k_{1}\},\{k_{2}\}
\begin{eqnarray}\label{}
S([k_{1}],[k_{2}]) &=& \xi\left(\frac{\beta}{2\pi}\right)^{4} \int\frac{d^4 \vec p}{(2\pi)^4}\frac{3\tanh \left(\pi\xi_p\right)-\pi\xi_p \left[2\pi\xi_p \tanh\left(\pi\xi_p\right)+3\right]\text{sech}^2\left(\pi\xi_p\right)}{4\xi_p^5}.
\end{eqnarray}
The mixed limits are given by
\begin{eqnarray}\label{}
S(\{k_{1}\},[k_{2}]) &=& S([k_{1}],\{k_{2}\}) = \xi\left(\frac{\beta}{2\pi}\right)^{4} \int\frac{d^4 \vec p}{(2\pi)^4} \frac{3\tanh\left(\pi\xi_p\right)-\pi\xi_p\text{sech}^2\left(\pi\xi_p\right)}{4\xi_{p}^{5}},
\end{eqnarray}
and the double long wavelength limit is written as
\begin{eqnarray}\label{}
S(\{k_{1}\},\{k_{2}\}) &=& \xi\left(\frac{\beta}{2\pi}\right)^{4} \int\frac{d^4 \vec p}{(2\pi)^4} \frac{3\tanh\left(\pi\xi_p\right)}{4\xi_{p}^{5}}.
\end{eqnarray}

In order to evaluate the loop integrals in the above coefficients, we use spherical coordinates in $4$-dimensions and consider the isotropy of the momentum space. The angular integral yields the solid angle, $2\pi^{2}$, while a change of variable in radial integral from $|\vec{p}|$ to $\zeta = \frac{\beta}{2\pi}\sqrt{|\vec{p}|^2+m^2}$ is performed, yielding for the double static limit,
\begin{eqnarray}\label{SL}
S([k_{1}],[k_{2}]) &=& \int_{|\xi|}^{\infty}d\zeta \frac{\xi\left(\zeta^2-\xi^2\right) \text{sech}^2(\pi\zeta) \left\{3\sinh(2\pi\zeta)-2\pi\zeta[2\pi\zeta\tanh (\pi\zeta)+3]\right\}}{64\pi^2\zeta^4}\nonumber\\
&=& \frac{\tanh(\pi\xi)}{16\pi^2} \equiv F(\xi),
\end{eqnarray}
for the mixed limits,
\begin{eqnarray}\label{}
S(\{k_{1}\},[k_{2}]) &=& \int_{|\xi|}^{\infty}d\zeta\frac{\xi\left(\zeta^2-\xi^2\right)\left[3\tanh(\pi\zeta)-\pi\zeta\text{sech}^2(\pi\zeta)\right]}{32\pi^2\zeta^4} \equiv G(\xi),
\end{eqnarray}
and for the double long wavelength limit,
\begin{eqnarray}\label{}
S(\{k_{1}\},\{k_{2}\}) &=& \int_{|\xi|}^{\infty}d\zeta\frac{3\xi\left(\zeta^2-\xi^2\right) \tanh(\pi\zeta)}{32\pi^2\zeta^4} \equiv H(\xi).
\end{eqnarray}

The result obtained in equation~(\ref{SL}) correspond to that one in equation~(\ref{EDT0}), showing that the derivative expansion method is consistent only to obtain the static limit. This fact is also found in three- \cite{Babu:1987, KaoYang:1993} and four-dimensional \cite{Nascimento:2000, Assuncao:2016fko} theories.

\subsection{Box Diagram}\label{Quadri}

Analogously, we analyze the nonanalyticity arising in the contribution to the CS term from the four-point function at finite temperature regime. In this case, in the evaluation of the trace, linear and cubic terms in the mass contribute to the CS term. Thus, considering only these terms, we have $\Pi^{\mu\nu\lambda\rho} \rightarrow \Pi_{CS}^{\mu\nu\lambda\rho}(k_{1},k_{2},k_{3})$, where
\begin{eqnarray}\label{C4}
\Pi_{{\footnotesize CS}}^{\mu\nu\lambda\rho} = ie^{4} \frac{i}{\beta}\sum_n\int\frac{d^4 \vec p}{(2\pi)^4}\frac{\left(G_{1}+k_{1}\cdot k_{2}\right) k_{3\alpha}-(k_{1}\cdot k_{3}) k_{2\alpha}+\left(G_{12}+k_{2}\cdot k_{3}\right) k_{1\alpha}}{\left(p^2+m^2\right) \left(p_{1}^{2}+m^2\right) \left(p_{12}^2+m^2\right) \left(p_{123}^2+m^2\right)} 4im \epsilon^{\mu\nu\lambda\rho\alpha},
\end{eqnarray}
with $p_{1}^{\mu}=p^{\mu}-k_{1}^{\mu}$, $p_{12}^{\mu}=p^{\mu}-k_{12}^{\mu}$, $p_{123}^{\mu}=p^{\mu}-k_{123}^{\mu}$ and $G_{i}=p_{i}^{2}+m^{2}$. However, in order to single out the nonanalytic contribution, we can use the coefficient-function $S(k_i,k_j)$, defined in the equations~(\ref{C1}) and (\ref{S}), to rewritten the equation~({\ref{C4}}) as follows:
\begin{eqnarray}\label{c4}
\Pi_{{\footnotesize CS}}^{\mu\nu\lambda\rho} &=& e^4\, S(k_1,k_{23})\,\epsilon^{\mu\nu\lambda\rho\alpha}k_{1\alpha}+e^4\, S(k_{12},k_{3})\,\epsilon^{\mu\nu\lambda\rho\alpha}k_{3\alpha}\nonumber\\
&+& \frac{4i^{3}m e^{4}}{\beta}\sum_n\int\frac{d^4 \vec p}{(2\pi)^4}\frac{(k_{1}\cdot k_{2}) k_{3\alpha}-(k_{1}\cdot k_{3}) k_{2\alpha}+(k_{2}\cdot k_{3}) k_{1\alpha}}{\left(p^2+m^2\right) \left(p_{1}^{2}+m^2\right) \left(p_{12}^2+m^2\right) \left(p_{123}^2+m^2\right)}\epsilon^{\mu\nu\lambda\rho\alpha}.
\end{eqnarray}
As we can see in the above equation~(\ref{c4}), the nonanalytic contributions are presented inside the coefficients $S(k_1,k_{23})$ and $S(k_{12},k_{3})$.  Also, there are analytic (remaining) contributions which are proportional to the external momenta $k_{i}^{\mu}$, in a way that if we choose any one of the limits $[k_i]=(k_{i0}=0,\vec k_{i}\rightarrow 0)$ or $\{k_i\}=(k_{i0}\rightarrow 0,\vec k_{i}=0)$, they yield a zero result.

As before, once the topological structure of the internal momentum is extracted, it follows the analysis over the coefficients $S(k_1,k_{23})$ and $S(k_{12},k_{3})$. It is possible to remove an external momentum, e.g., $k_{2}^{\mu}\rightarrow 0$, so that $S(k_1,k_{23})\rightarrow S(k_1,k_{3})$ and $S(k_{12},k_{3})\rightarrow S(k_1,k_{3})$. This freedom follows because the nonanalytic behavior of the function $S(k_{i},k_{j})$ around the origin was already evaluated in previous subsection for $k_i=k_1$ and $k_j=k_{2}$. Hence, we have
\begin{eqnarray}\label{}
\Pi_{{\footnotesize CS}}^{\mu\nu\lambda\rho}(k_1,0,k_3) = e^4\, S(k_1,k_3)\,\epsilon^{\mu\nu\lambda\rho\alpha}k_{13\alpha},
\end{eqnarray}
where $k_{13\alpha}=k_{1\alpha}+k_{3\alpha}$. Thus, the limits about external momenta, through four distinct paths, are given by $S([k_1],[k_3])$  equal to $F(\xi)$, in double static limit, $S(\{k_1\},[k_3])=S([k_1],\{k_3\})$ equal to $G(\xi)$, in the mixed limit, and $S(\{k_1\},\{k_3\})$ equal to $H(\xi)$, in double long wavelength limit.

\subsection{Pentagon Diagram}\label{Pen}

Finally, we proceed with the same analysis for the contribution to the CS term arising from the five-point function. At this time, in evaluating the trace, linear, cubic, and quintic terms in the mass contribute to the CS term, through to the convergent integral $\Pi^{\mu\nu\lambda\rho\sigma} \rightarrow \Pi_{CS}^{\mu\nu\lambda\rho\sigma}(k_{1},k_{2},k_{3},k_{4})$, with
\begin{eqnarray}\label{C5}
\Pi_{{\footnotesize CS}}^{\mu\nu\lambda\rho\sigma} &=& ie^{5} \frac{i}{\beta}\sum_n\int\frac{d^4 \vec p}{(2\pi)^4}\frac{-4im\epsilon^{\mu\nu\lambda\rho\sigma}}{\left(p^2+m^2\right) \left(p_{1}^{2}+m^2\right) \left(p_{12}^2+m^2\right) \left(p_{123}^2+m^2\right) \left(p_{1234}^2+m^2\right)}\nonumber\\
&\times& \left[\left(G_1+k_1\cdot k_2\right) \left(G_{123}+k_3\cdot k_4\right)+\left(G_{12}+k_2\cdot k_3\right)(k_1\cdot k_4)-(k_1\cdot k_3) (k_2\cdot k_4)\right],
\end{eqnarray}
where $p_{1}^{\mu}=p^{\mu}-k_{1}^{\mu}$, $p_{12}^{\mu}=p^{\mu}-k_{12}^{\mu}$, $p_{123}^{\mu}=p^{\mu}-k_{123}^{\mu}$ and $G_{i}=p_{i}^{2}+m^{2}$. Again, we found analytic contributions which are proportional to the external momenta $k_{i}^{\mu}$ and vanish in any one of the limits $[k_{i}]=(k_{i0}=0,\vec k_{i}\rightarrow 0)$ or $\{k_{i}\}=(k_{i0}\rightarrow 0,\vec k_{i}=0)$, so that only the nonanalytic contribution proportional to $G_{1}G_{12}$ yields the coefficient-function $S(k_{12},k_{1234})$. Then, after removing, e.g., the momenta $k_{2}^{\mu}$ and $k_{4}^{\mu}$ and taking the limits, we obtain
\begin{eqnarray}\label{}
\Pi_{{\footnotesize CS}}^{\mu\nu\lambda\rho\sigma}(k_1,0,k_3,0) &=& -e^5\, S(k_1,k_{13})\,\epsilon^{\mu\nu\lambda\rho\sigma},
\end{eqnarray}
where newly $S([k_1],[k_{13}])=F(\xi)$, in double static limit, $S(\{k_1\},[k_{13}])=S([k_1],\{k_{13}\})=G(\xi)$, in the mixed limit, and $S(\{k_1\},\{k_{13}\})=H(\xi)$, in double long wavelength limit.

\subsection{The Five-Dimensional Nonabelian Chern-Simons Term at Finite Temperature}\label{FT}

The one-loop induced nonabelian five-dimensional CS action is obtained by adding the odd-contribution from the triangle, box, and pentagon graphs as follows:
\begin{equation}\label{CSAction}
S_{CS}=\frac{S(0,0)}{3}\int d^5x \epsilon^{\mu\nu\lambda\rho\sigma}[-e^3(\partial_{\nu}A_{\mu})(\partial_{\rho}A_{\lambda})A_{\sigma}+ie^4\frac{3}{2}(\partial_{\sigma}A_{\mu})A_{\nu}A_{\lambda}A_{\rho}+e^5\frac{3}{5}A_{\mu}A_{\nu}A_{\lambda}A_{\rho}A_{\sigma}],
\end{equation}
where, $S(0,0)$, as in the previous subsections, is equal to $F(\xi)$, $G(\xi)$, or $H(\xi)$ in the double static limit, mixed limit, or double long wavelength limit, respectively.

As expected, at a finite temperature regime, the one-loop five-dimensional CS action is a nonanalytic but covariant function, so that the thermal effects acts only on the coefficient. At the same time, the topological structure is preserved without any change. This fact was also observed in three-dimensional CS action \cite{KaoYang:1993}, and it is expected to hold, in general, in odd-dimension once the topological structure characteristic of CS term is obtained yet at integrand level.

\begin{figure}[!h]
\begin{center}\includegraphics[scale=0.5]{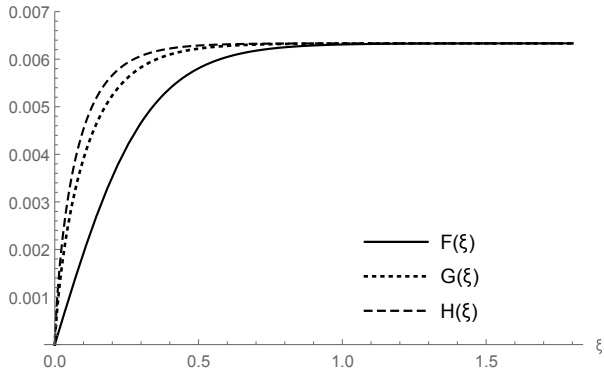}
\caption{Plot of functions $F(\xi)$, $G(\xi)$ and $H(\xi)$.}\label{fig2}
\end{center}
\end{figure}

We also discuss the temperature dependence for the coefficient of the CS term. The behavior of $F(\xi)$, $G(\xi)$, and $H(\xi)$ are numerically plotted in Fig~\ref{fig2}.

We observe that at zero and infinite temperature, the limits coincide, and then $\Pi_{{\footnotesize CS}}^{\mu\nu\lambda}$ recover the analyticity. When $T\rightarrow 0$ $(\xi\rightarrow \infty)$ the result obtained by \cite{Grimm} is found. Alternatively, the limit $T\rightarrow \infty$ ($\xi\rightarrow 0$ or $m\rightarrow 0$) vanishes. This result is in agreement with the fact, well-known in the literature, that the fermionic mass is a parity-odd quantity in odd-dimension responsible for the induction of CS term. Between these extremes, as the temperature increases, the fermion mass ($\xi = \beta m/2\pi$) is attenuated by thermal effects until completely suppressed at $T\rightarrow \infty$.

\section{Summary}

As exposed above, we have studied the nonanalytical behavior of the CS coefficient at finite temperature. Firstly, we have shown that we can single out from the three graphs, through derivative expansion, the same coefficient that generates the nonabelian five-dimensional CS term, namely, the triangle, box, and pentagon ones depicted in Fig~\ref{fig1}. We have then calculated its finite temperature contribution solving the integral before the sum over the Matsubara frequencies. Our result is in agreement with the results found in \cite{Grimm, Witten}. 

We have verified the nonanalytical behavior of the coefficient through nonperturbative calculations of the triangle, box, and pentagon diagrams. These three diagrams were calculated individually and they gave rise to the same result for the three limits, namely, the double static limit, the mixed limit, and the double wavelength limit. We also have observed that the topological structure remains unchanged even under thermal effects.

\begin{acknowledgments}
This work was supported by Conselho Nacional de Desenvolvimento Cient\'{i}fico e Tecnol\'{o}gico (CNPq) and Coordena\c{c}\~ao de Aperfei\c{c}oamento de Pessoal de N\'ivel Superior (Capes).
\end{acknowledgments}

\end{document}